\documentclass[]{spie}  
\usepackage[]{graphicx}
\usepackage{amsmath}
\usepackage{enumerate}
\usepackage{color}
\usepackage{cite}
\usepackage{mathptmx}      
\usepackage{bm}

\title{Data analysis of Einstein-Podolsky-Rosen-Bohm laboratory experiments}

\author{H. De Raedt\supit{a}, F. Jin\supit{b}, and K. Michielsen\supit{b,}\supit{c}
\skiplinehalf
\supit{a}
Department of Applied Physics,
Zernike Institute for Advanced Materials,
University of Groningen, Nijenborgh 4, NL-9747 AG Groningen, The Netherlands
\\
\supit{b}
Institute for Advanced Simulation, J\"ulich Supercomputing Centre,
Forschungszentrum J\"ulich, D-52425 J\"ulich, Germany
\\
\supit{c}
RWTH Aachen University, D-52056 Aachen, Germany
}

\authorinfo{
Further author information: (Send correspondence to H. De Raedt)\\
H. De Raedt : E-mail: h.a.de.raedt@rug.nl \\
F. Jin: E-mail: F.Jin@fz-juelich.de \\
K. Michielsen: E-mail: k.michielsen@fz-juelich.de}

\newcommand{\url}[1]{{\rm #1}}

\newcommand\ns{\,\mathrm{ns}}
\newcommand\mus{\,\mu\mathrm{s}}

\begin{document}
  \maketitle

\begin{abstract}
Data sets produced by three different Einstein-Podolsky-Rosen-Bohm (EPRB) experiments
are tested against the hypothesis that the statistics of this data is described by quantum theory.
Although these experiments generate data that violate Bell inequalities
for suitable choices of the time-coincidence window,
the analysis shows that it is highly unlikely that these data sets are compatible with
the quantum theoretical description of the EPRB experiment, suggesting that
the popular statements that EPRB experiments agree with quantum theory lack
a solid scientific basis and that more precise experiments are called for.
\end{abstract}



\section{Introduction}\label{introduction}

In scientific and popular literature on quantum physics, it is quite common to find statements that
the experimental results~\cite{KOCH67,FREE72,CLAU78,ASPE82b,TAPS94,TITT98,WEIH98,WEIH00,WEIH07,ROWE01,FATA04,SAKA06,HNIL07}
of Einstein-Podolsky-Rosen-Bohm (EPRB) experiments are described by quantum theory~\cite{CLAU78,BELL93,BALL03}.
While it is firmly established that the experimental data show violations of a Bell inequality,
it is remarkable that this observation alone is taken to imply that the experimental
data gathered in EPRB experiments is indeed complying
with the predictions of quantum theory for this particular experiment.
Indeed, as shown below, the statistical variation of the data produced by EPRB experiments
seems to be much larger than for most experiments which made quantum theory famous.
For instance, the energy of photons emitted from atoms is reproducible to many
digits and theory and experiments agree very well but
the data taken in EPRB experiments deviate significantly (more than 4-5 standard deviations)
from the quantum theoretical predictions.

The data sets generated by the experiment of Weihs {\sl et al.} have been
scrutinized earlier~\cite{HNIL02,HNIL07,ADEN07,BIGE09,AGUE09,BIGE11,RAED12}
but, with the exception of Ref.~\citen{RAED12} there is no report which scrutinizes the idea
that the data gathered in EPRB experiments passes the
hypothesis test that it complies with the quantum theoretical
description of the EPRB thought experiment, that is with a description
in terms of a state which does not depend on the setting of the analyzers.

One reason for deeming such a test unnecessary may be rooted in the widespread
misconception that Bell~\cite{BELL64,BELL93} has proven that a violation of one of his inequalities
implies that the experimental findings rule out {\bf any} explanation
in terms of {\it all} classical (Hamiltonian as well as non-Hamiltonian) models
that satisfy Einstein's criteria for local causality.
Although the general validity of Bell's result has been questioned by many workers~\cite{%
PENA72,FINE74,FINE82,FINE82a,FINE82b,MUYN86,KUPC86,JAYN89,%
BROD93,PITO94,FINE96,KHRE99,SICA99,BAER99,%
HESS01,HESS05,ACCA05,KRAC05,SANT05,LOUB05,%
KUPC05,MORG06,KHRE07,ADEN07,NIEU09,MATZ09,KARL09,KHRE09,GRAF09,KHRE11,NIEU11,RAED11a},
even if Bell's results were generally applicable,
it still remains to be shown that the experimental data complies with the predictions of quantum theory.

In this paper, we report results of the hypothesis test applied to data produced
by the EPRB experiments of Weihs {\sl et al.}~\cite{WEIH98,WEIH00},
M.B. Ag\"uero {\sl et al.}~\cite{AGUE09}
and G. Adenier {\sl et al.}~\cite{ADEN12,VIST12}.
We present compelling evidence that the experimental data,
while violating Bell inequalities, does not support the hypothesis that the data complies
with the predictions of quantum theory for the EPRB thought experiment,
suggesting that in the real experiments there are processes at work
which deserve to be identified and studied further.

\section{Hypothesis test}\label{test}

According to quantum theory of the EPRB thought experiment, the results of repeated measurements
of the system of two $S=1/2$ particles in the spin state $|\Phi\rangle$
are given by the single-spin expectation values
\begin{eqnarray}
\widehat E_1(\mathbf{a})&=&\langle \Phi|\mathbf{S}_1\cdot\mathbf{a}|\Phi\rangle=\langle \Phi|\mathbf{S}_1|\Phi\rangle\cdot\mathbf{a}
,
\nonumber \\
\widehat E_2(\mathbf{b})&=&\langle \Phi|\mathbf{S}_2\cdot\mathbf{b}|\Phi\rangle=\langle \Phi|\mathbf{S}_2|\Phi\rangle\cdot\mathbf{b}
,
\label{Ei}
\end{eqnarray}
and the two-particle correlations
%
$
\widehat E(\mathbf{a},\mathbf{b})=
\langle \Phi|\mathbf{S}_1\cdot\mathbf{a}\; \mathbf{S}_2\cdot\mathbf{b}|\Phi \rangle
=\mathbf{a}\cdot\langle \Phi|\mathbf{S}_1\; \mathbf{S}_2|\Phi \rangle \cdot\mathbf{b}
$
where $\mathbf{a}$ and $\mathbf{b}$ specify the directions of the analyzers.
We have introduced the notation $\widehat{\phantom{E}}$
to distinguish the quantum theoretical prediction from the results
obtained by analysis of the experimental data sets.

Quantum theory of the EPRB thought experiment assumes that $|\Phi\rangle$ does
not depend on $\mathbf{a}$ or $\mathbf{b}$.
Therefore, from Eq.~(\ref{Ei}) it follows immediately that
$\widehat E_1(\mathbf{a})$ does not depend on $\mathbf{b}$
and that
$\widehat E_2(\mathbf{b})$ does not depend on $\mathbf{a}$.
Under the hypothesis that quantum theory describes the
data collected in the laboratory EPRB experiment,
we may expect that
$E_1(\mathbf{a},\mathbf{b})\approx\widehat E_1(\mathbf{a})$
and
$E_2(\mathbf{a},\mathbf{b})\approx\widehat E_2(\mathbf{b})$,
exhibit the same independencies.
This is the basis of our test.

In practice, we are dealing with real data and not with mathematical expressions
such as Eq.~(\ref{Ei})
and therefore, we need a criterion to decide whether or not the data
is in agreement with the value predicted by quantum theory of the EPRB experiment.
In line with standard statistical hypothesis testing, we adopt the following criterion:
\begin{center}
\framebox{
\parbox[t]{0.92\hsize}{%
The data for $E_1(\mathbf{a},\mathbf{b})$ ($E_2(\mathbf{a},\mathbf{b})$) are considered to be in conflict with
the prediction of quantum theory of the EPRB thought experiment if it shows a dependency
on $\mathbf{b}$ ($\mathbf{a}$)
that exceeds five times the upper bound $1/\sqrt{N_c}$ to the standard deviation $\sigma_{N_c}$.
}
}
\end{center}

We emphasize that data that does not satisfy our criterion exhibits a spurious kind of ``non-locality''
that {\bf cannot} be explained by the quantum theoretical description of the EPRB experiment.
A key feature of this hypothesis test is that it does not rely on any particular property
of the state $|\Phi\rangle$.
For instance, if in a laboratory EPRB experiment we find that
$E_1(\mathbf{a},\mathbf{b})$ shows a dependence on $\mathbf{b}$ that exceeds five times the standard deviation,
this dependence {\bf cannot} be attributed to
$|\Phi\rangle$ deviating from the singlet state.

\begin{figure}[t]
\begin{center}
\includegraphics[width=14cm]{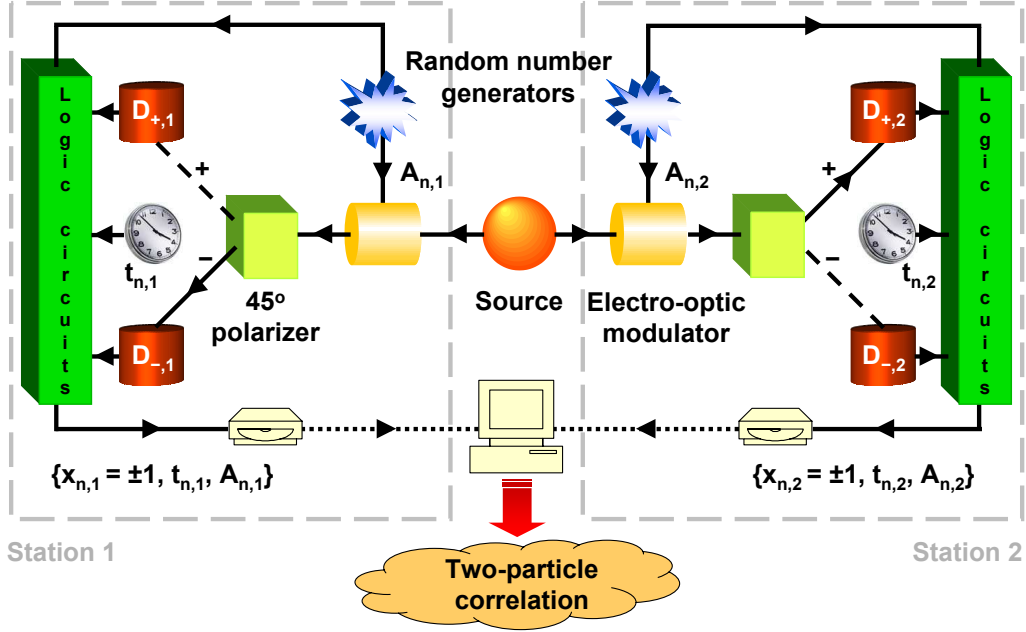}
\caption{Schematic diagram of an EPRB experiment with photons performed by Weihs {\sl et al.}~\cite{WEIH98,WEIH00}
A light source emits pairs of photons in spatially distinct directions
As a photon arrives at station $i=1,2$ it passes through an electro-optic
modulator (EOM) which rotates the polarization of the photon that passes through it
by an angle $\theta_i$ which is determined by the voltage applied to the EOM.
The latter is controlled by a binary variable $A_i$, which is chosen at random~\cite{WEIH98,WEIH00}.
As the photon leaves the EOM, a polarizing beam splitter directs the photon to one of the two detectors,
may produce a signal $x_{n,i}=\pm1$ where the subscript $n$ labels the $n$th detection event.
For each click of a detector, a clock at station $i$ generates a time tag $t_{n,i}$.
}
\label{fig:Weihs}
\end{center}
\end{figure}

In the three experiments that we consider in this paper, the polarization of each photon is used
as the spin-1/2 degree-of-freedom in Bohm's version~\cite{BOHM51}
of the EPR gedanken experiment~\cite{EPR35}.
In these experiments the unit vectors $\mathbf{a}$ or $\mathbf{b}$ are assumed to be coplanar
and are specified by the angles $a$ and $b$.
All experiments use light sources that may emit pairs of photons.
Note that the statistical properties of a large set of these photons are a priori unknown
and have to be inferred from the analysis of the recorded detection events.
The popular statement that such sources are emitting singlets only means that
the source is emitting pairs of photons, the correlation of which may be described
by the quantum theory of a pair of two spin-1/2 objects.

\section{EPRB laboratory experiment I}\label{facts}

The schematic diagram of the first EPRB experiment with photons that we analyze is
shown in Fig.~\ref{fig:Weihs}.
As a photon arrives at station $i=1,2$ ($i=1,2$ denotes Alice's and Bob's station, respectively),
it passes some optical components and may generate a click in one of the detectors.
We say ``may'' because in general the detection efficiency is rather low, of the order of a few percent.
The firing of a detector defines an event.
At the $n$th event at station $i$,
the dichotomic variable $A_{n,i}$ (see Fig.~\ref{fig:Weihs}),
controlling the rotation angle $\theta_{n,i}$,
the dichotomic variable $x_{n,i}$ designating which detector fires,
and the time tag $t_{n,i}$ of the detection event
are written to a file on a hard disk,
allowing the data to be analyzed long after the experiment has terminated~\cite{WEIH98,WEIH00}.
The set of data collected at station $i$ may be written as
\begin{eqnarray}
\label{Ups}
\Upsilon_i=\left\{ {x_{n,i},t_{n,i},\theta_{n,i} \vert n =1,\ldots ,N_i } \right\}
,
\end{eqnarray}
where we allow for the possibility that the number of detected events $N_i$
at stations $i=1,2$ need not (and in practice is not) to be the same
and we have used the rotation angle $\theta_{n,i}$ instead
of the corresponding dichotomic variable $A_{n,i}$ to facilitate the
comparison with the quantum theoretical description.
The data sets $\{\Upsilon_1,\Upsilon_2\}$, provided to us by G. Weihs,
are the starting point for the analysis presented in this paper.

\subsection{Data processing procedure}\label{processing}

In general, it is unknown when the source will emit a pair or just a single photon.
In fact, the only thing that can be said with certainty is that
one or more detectors produced a click, all other statements being
inferences based on the observed clicks.
Therefore, to establish that the source has emitted a pair of photons,
it is necessary to introduce a criterion to distinguish single-photon
from two-photon events.
In EPRB experiments with photons, this classification is made
on the basis of coincidence in time~\cite{WEIH98,CLAU74}.
Adopting the procedure employed by Weihs {\sl et al.}~\cite{WEIH98,WEIH00},
coincidences are identified by comparing the time differences
$t_{n,1}-t_{m,2}$ with a window $W$~\cite{WEIH98,WEIH00,CLAU74,AGUE09},
where $n=1,\ldots,N_1$ and $m=1,\ldots,N_2$.

By definition, for each pair of rotation angles $a$ and $b$,
the number of coincidences between detectors $D_{x,1}$ ($x =\pm $1) at station 1 and
detectors $D_{y,2}$ ($y =\pm $1) at station 2 is given by
\begin{eqnarray}
\label{Cxy}
C_{xy}=C_{xy}(a,b)&=&
\sum_{n=1}^{N_1}
\sum_{m=1}^{N_2}
\delta_{x,x_{n ,1}} \delta_{y,x_{m ,2}}
\delta_{a ,\theta_{n,1}}\delta_{b,\theta_{m,2}}
\Theta(W-\vert t_{n,1} -t_{m ,2}\vert)
,
\end{eqnarray}
where $\Theta (t)$ is the Heaviside step function.
In Eq.~(\ref{Cxy}) the sum over all events has to be carried out such that each event (= one detected photon) contributes only once.
Clearly, this constraint introduces some ambiguity in the counting procedure as there is a priori, no clear-cut criterion
to decide which events at stations $i=1$ and $i=2$ should be paired.
One obvious criterion might be to choose the pairs such that $C_{xy}$ is maximum but as we explain later,
such a criterion renders the data analysis procedure (not the data production!) acausal.
It is trivial though (see later) to analyse the data generated by the experiment of Weihs {\sl et al.}
such that conclusions do not suffer from this artifact.
In general, the values for the coincidences
$C_{xy}(a,b)$ depend on the time-tag resolution $\tau$
and the window $W$ used to identify the coincidences.

The single-particle averages and correlation between the coincidence counts
are defined by
\begin{eqnarray}
\label{Exy}
E_1(a,b)&=&
\frac{\sum_{x,y=\pm1} x C_{xy}}{\sum_{x,y=\pm1} C_{xy}}
=
\frac{C_{++}-C_{--}+C_{+-}-C_{-+}}{C_{++}+C_{--}+C_{+-}+C_{-+}}
\nonumber \\
E_2(a,b)&=&
\frac{\sum_{x,y=\pm1} yC_{xy}}{\sum_{x,y=\pm1} C_{xy}}
=
\frac{C_{++}-C_{--}-C_{+-}+C_{-+}}{C_{++}+C_{--}+C_{+-}+C_{-+}}
\nonumber \\
E(a,b)&=&
\frac{\sum_{x,y=\pm1} xy C_{xy}}{\sum_{x,y=\pm1} C_{xy}}
=
\frac{C_{++}+C_{--}-C_{+-}-C_{-+}}{C_{++}+C_{--}+C_{+-}+C_{-+}}
,
\end{eqnarray}
where the denominator $N_c=C_{++}+C_{--}+C_{+-}+C_{-+}$
in Eq.~(\ref{Exy}) is the sum of all coincidences.
In practice, coincidences are determined by a four-step procedure~\cite{WEIH00}:
\begin{enumerate}
\item{Compute a histogram of time-tag differences $t_{n,1}-t_{m,2}$ of pairs of detection events.}
\item{Determine the time difference $\Delta_G$ for which this histogram shows a maximum.}
\item{Add $\Delta_G$ to the time-tag data of Alice, thereby moving the position of the maximum of the histogram to zero.}
\item{Determine the coincidences using the new time-tag differences,
each photon contributing to the coincidence count at most once.}
\end{enumerate}
The global offset, denoted by $\Delta_G$, may be attributed
to the loss of synchronization of the clocks used in the stations of Alice and Bob~\cite{WEIH00}.

However, the use of a global offset $\Delta_G$, determined
by maximizing the number of coincidences, introduces an element of non-causality in the
analysis of the experimental results (but not in the original data itself).
Whether or not at a certain time, a pair contributes to the number of coincidences depends
on {\bf all} the coincidences, also on those at later times.
This is an example of an acausal filtering~\cite{PRES03}.
The output (coincidence or not) depends on both all previous
and all later inputs (the detection events and corresponding time-tags).
Summarizing:

\begin{center}
\framebox{
\parbox[t]{0.92\hsize}{%
Employing a global offset $\Delta_G$ to maximize the number of coincidences
makes the results of EPRB experiments explicitly acausal and renders
considerations about coincidences happening inside or outside the light cone irrelevant.
}
}
\end{center}

As it is our aim to test whether the data of the EPRB experiments comply with quantum theory,
not to find the maximum violation of some inequality,
we will not dwell on this issue any further and
simply discard conclusions that depend on the use of a non-zero $\Delta_G$.

\begin{figure}[t]
\centering
\includegraphics[width=7.5cm]{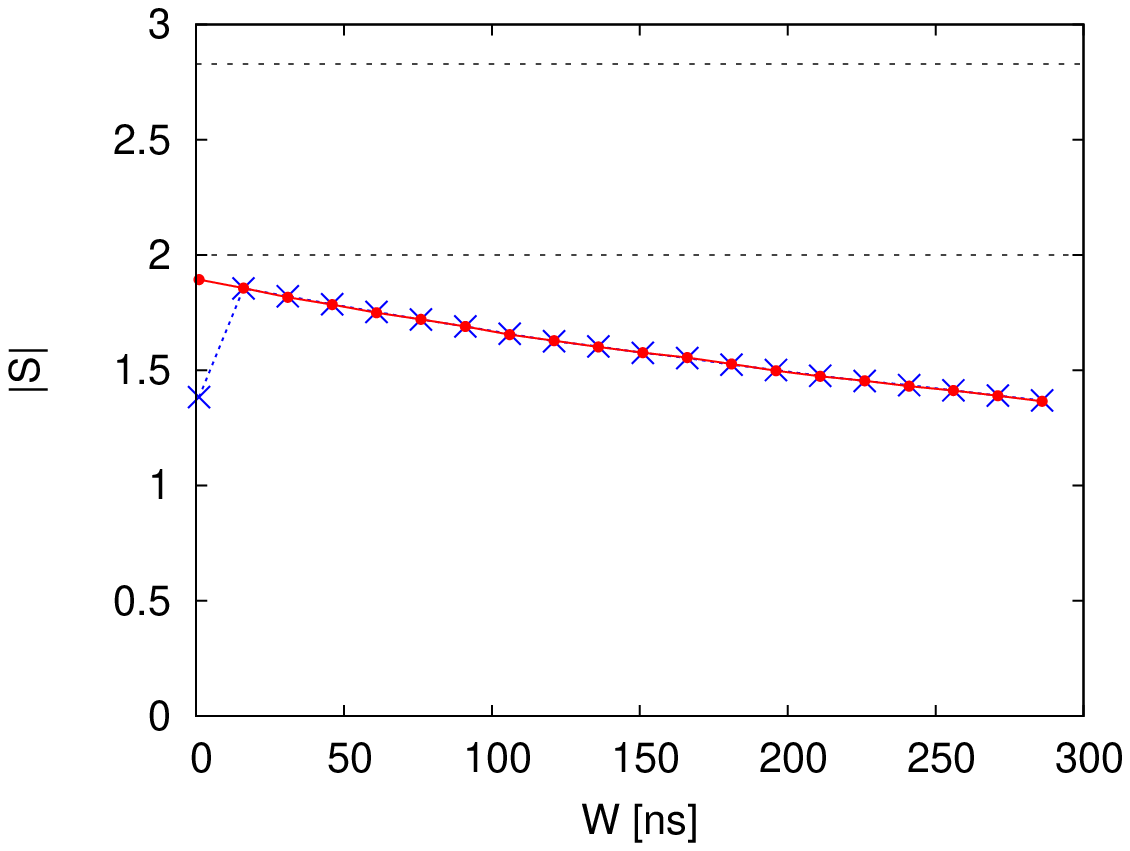}
\includegraphics[width=7.5cm]{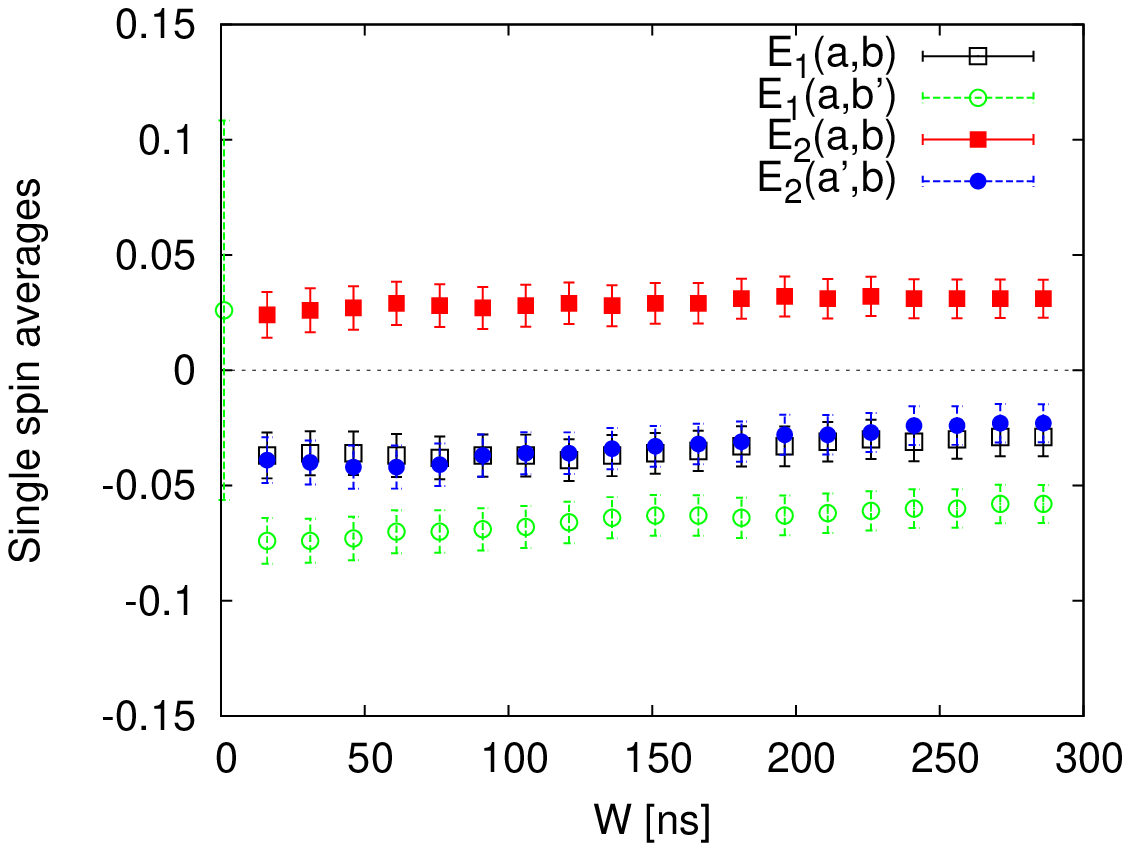}
\includegraphics[width=7.5cm]{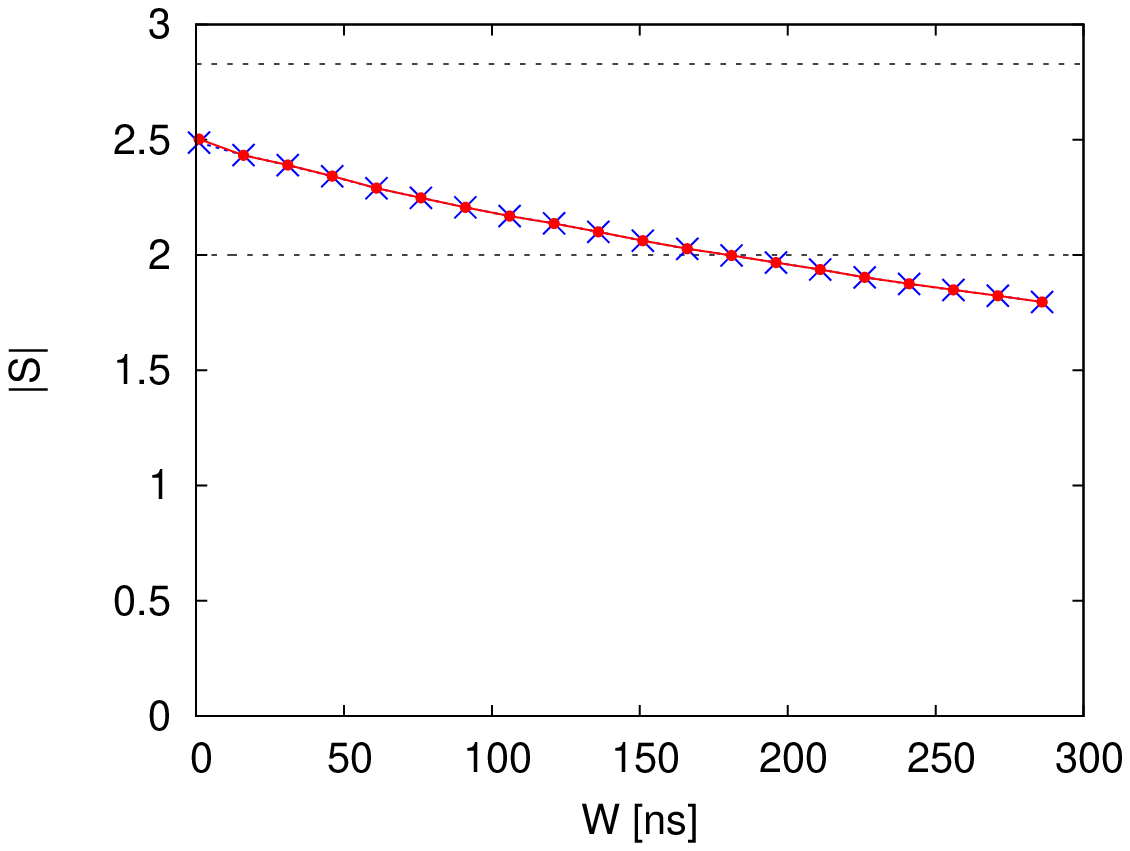}
\includegraphics[width=7.5cm]{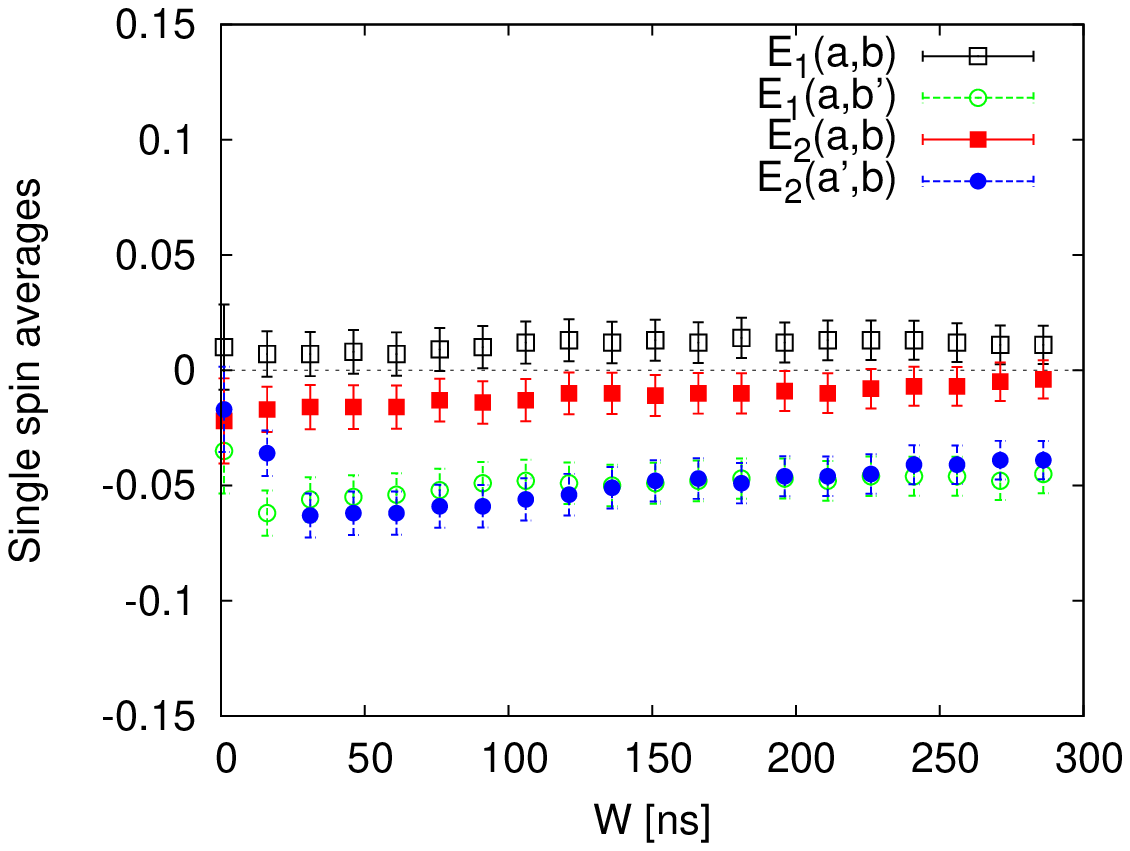}
\caption{
Analysis of the data sets {\bf newlongtime1} (top) and {\bf newlongtime2} (bottom),
recorded in experiments performed by Weihs {\sl et al.}~\cite{WEIH98,WEIH00}
Left:
$|S|=|E(a,b)-E(a,b')+E(a',b)+E(a',b')|$ as a function of the time window $W$ and
$a=0$, $a'=\pi/4$, $b=\pi/8$ and $b'=3\pi/8$.
The dashed lines represent the maximum value for a quantum system of two $S=1/2$ particles
in a separable (product) state ($|S|=2$) and in a singlet state ($|S|=2\sqrt{2}$), respectively.
If $|S|>2$, a Bell inequality is violated.
Blue crosses: $\Delta_G=0$.
Red bullets connected by the red solid line: $\Delta_G=0.5\ns$.
Note that the data {\bf newlongtime1} do not lead to violations of the Bell inequality
whereas the data {\bf newlongtime2} do.
Right:
Selected single-particle averages
as a function of $W$ and for $\Delta_G=0$.
Error bars correspond to $2.5$ standard deviations.
For small $W$, the total number of coincidences
is too small to yield statistically meaningful results.
For $W>20\ns$ the change of some of these single-spin averages
observed by Bob (Alice) when Alice (Bob) changes her (his) setting,
systematically exceeds five standard deviations,
suggesting it is highly unlikely that the data is in
concert with quantum theory of the EPRB experiment.
}
\label{fig.A5}
\end{figure}

\subsection{Data analysis: Results}\label{results}

Figure~\ref{fig.A5}(left) shows the typical results of the Bell function
$S=E(a,b)-E(a,b')+E(a',b)+E(a',b')$ as a function of $W$
obtained by processing two different data sets ({\bf newlongtime1} and {\bf newlongtime2})
in the collection provided by G. Weihs.
Clearly, the data {\bf newlongtime1} does not show any violation of a Bell inequality at all
but {\bf newlongtime2} does because for $W<150\ns$, the Bell inequality $|S|\le2$ is clearly violated.
For $W>200\ns$, much less than the average time ($>30\mus$) between two coincidences,
the Bell inequality $|S|\le2$ is satisfied, demonstrating that the ``nature''
of the emitted pairs is not an intrinsic property of the pairs themselves but also depends on
the choice of $W$ made by the experimenter.
For $W>20\ns$, there is no significant statistical evidence that the ``noise'' on the data depends on $W$
but if the only goal is to maximize $|S|$, it is expedient to consider $W<20\ns$.
From these experimental results, it is clear that the time-coincidence window does not constitute a ``loophole''.
Not only is it an essential element of these EPRB experiments,
it also acts as a filter that is essential for the data to violate a Bell inequality.

Figure~\ref{fig.A5}(right) shows results of a selection of single-particle expectations as a function of $W$
extracted from both {\bf newlongtime1} (top) and {\bf newlongtime2} (bottom).
It is clear that these results violate our criterion for being compatible
with the prediction of quantum theory of the EPRB model.
Indeed, we find that $E_1(a,b)\not=E_1(a,b')$ and $E_2(a,b)\not=E_2(a',b)$
which contradict the quantum theoretical result Eq.~(\ref{Ei}).
In other words, some of the single-particle averages measured by Alice (Bob)
show, beyond 5 standard deviations, correlations with the settings chosen
by Bob (Alice).
Therefore, according to standard practice of hypothesis testing, the likelihood that this data set
can be described by the quantum theory of the EPRB experiment should be considered as extremely small.
This finding is not an accident:
Our analysis of 23 data sets produced by the experiment of Weihs {\sl et al.}
shows that none of these data sets satisfies our hypothesis test for being compatible with
the predictions of quantum theory of the EPRB model.

\subsection{Role of the detection efficiency}\label{detection}

For the experimental set up of Weihs {\sl et al.},
the dependence of $E_1(a,b)$ ($E_2(a,b)$) on ${b}$ (${a}$) cannot
be attributed to detection efficiencies of the detectors at the stations.
This can be seen as follows.

As photons may be absorbed when passing through the EOM and
as detectors do not register all incident photons, we may write
\begin{eqnarray}
C_{xy}(a,b)&=&\kappa_1(a)\kappa_2(b)\eta_1(x)\eta_2(y)N P(xy|ab)
,
\end{eqnarray}
where $0<\kappa_i(.)\le1$ and $0<\eta_i(.)\le1$
represent the efficiency of the EOM and detectors at station $i=1,2$ respectively,
$N$ is the number of photon pairs emitted by the source
and $P(xy|ab)=(1+x\widehat E_1({a})+y\widehat E_2({b})+xy\widehat E({a},{b}))/4$
is the most general form of the probability for a pair $x,y=\pm1$, compatible with quantum theory of the EPRB thought experiment.
Note that the experiment of Weihs {\sl et al.} uses polarizing beam splitters, and therefore the
detectors receive photons with fixed polarization.
Hence the detection efficiencies $\eta_1(x)$ and $\eta_2(y)$ are not expected
to depend on the polarization of the photons that leave the EOMs~\cite{WEIH00}.
After some elementary algebra, we find
\begin{eqnarray}
E_1(a,b)&=&
\frac{r_1+\widehat E_1({a})+r_1r_2\widehat E_2({b})+r_2\widehat E({a},{b})}{
1+r_2\widehat E_1({a})+r_1\widehat E_2({b})+r_1r_2\widehat E({a},{b})}
.
\nonumber \\
E_2(a,b)&=&
\frac{r_2+r_1r_2\widehat E_1({a})+\widehat E_2({b})+r_1\widehat E({a},{b})}{
1+r_2\widehat E_1({a})+r_1\widehat E_2({b})+r_1r_2\widehat E({a},{b})}
,
\nonumber \\
E(a,b)&=&
\frac{r_1r_2+r_2\widehat E_1({a})+r_1\widehat E_2({b})+\widehat E({a},{b})}{
1+r_2\widehat E_1({a})+r_1\widehat E_2({b})+r_1r_2\widehat E({a},{b})}
,
\label{de0}
\end{eqnarray}
where $-1\le r_i=(\eta_i(+1)-\eta_i(-1))/(\eta_i(+1)+\eta_i(-1))\le 1$
parameterizes the relative efficiencies of the detectors at stations $i=1,2$.
The parameters $r_1$, $r_2$ and the state of the quantum system (fully described by
$\widehat E_1({a})$, $\widehat E_2({b})$, $\widehat E({a},{b})$)
can be determined by considering the set of nine equations for the
three pairs of settings.
Taking for instance the data for the set of settings $\{(a,b),(a,b'),(a',b)\}$,
yields three times three equations of the form Eq.~(\ref{de0})
with nine unknowns. These sets of equations are easily solved numerically.
As an illustrative example, we examine the data at $W=50\ns$ taken from the experimental data set
{\bf newlongtime2}, see also Fig.~\ref{fig.A5}.
The results of the numerical solution of the nine equations
is given by the first row of Table~\ref{tab:a}.
Similarly, we can construct three additional sets of nine equations.
Their solutions are also presented in Table~\ref{tab:a}.
Clearly, there is no way in which these four solutions can be considered as compatible.
Needless to say, the full set of 12 equations has no solution at all.
This incompatibility is not accidental, it is generic.
Using a different approach of analyzing the data, J.H. Bigelow
came to a similar conclusion~\cite{BIGE09}.
Apparently, including a model for the detector efficiency does not resolve the conflict
between the experimental data of Weihs {\sl et al.} and quantum theory of the EPRB thought experiment.

\begin{table}[t]
\begin{center}
\caption{Results of the numerical solution of the set of three groups of equations such as Eq.~(\ref{de0})
for data taken from the experimental data set
{\bf newlongtime2} at $W=50\ns$.
The pair of settings that has not been included corresponds to the entry indicated with the --.
For instance, the results of the first row have been obtained by excluding the data
for the setting $(a',b')$.
Consistency demands that the numbers in each column are close to each other.
The fact that the four solutions are inconsistent suggests that the data is incompatible
with quantum theory for the EPRB thought experiment.
}
\begin{tabular}{cccccccccc}
\noalign{\vskip 6pt}
\hline
\noalign{\vskip 4pt}
    {$r_1$}
  & {$r_2$}
  & {$\widehat E_1({a})$}
  & {$\widehat E_1({a'})$}
  & {$\widehat E_2({b})$}
  & {$\widehat E_2({b'})$}
  & {$\widehat E({a},{b})$}
  & {$\widehat E({a},{b'})$}
  & {$\widehat E({a'},{b})$}
  & {$\widehat E({a'},{b'})$}
  \\
\hline
\noalign{\vskip 4pt}
$+0.17$ & $-0.01$ & $-0.17$ & $-0.23$ & $+0.11$ & $-0.13$ & $-0.72$ & $+0.47$ & $-0.52$ & -- \\
$-0.06$ & $-0.01$ & $+0.06$ & $+0.03$ & $-0.05$ & $-0.03$ & $-0.69$ & $+0.45$ & --      & $-0.71$ \\
$+0.17$ & $-0.16$ & $-0.29$ & $-0.32$ & $+0.27$ & $+0.31$ & $-0.83$ & --      & $-0.62$ & $-0.87$ \\
$-0.06$ & $-0.17$ & $+0.13$ & $-0.08$ & $+0.16$ & $+0.13$ & --      & $+0.46$ & $-0.50$ & $-0.72$ \\
\hline
\end{tabular}
\label{tab:a}
\end{center}
\end{table}

\section{EPRB laboratory experiment II}\label{facts2}

Qualitatively, the data obtained from
the less sophisticated implementation of the EPRB experiment (compared to the experiment of Weihs {\sl et al.})
with photons by M.M. Ag\"uero {\sl et al.}~\cite{AGUE09}
exhibits similar features as the data produced by the experiment by Weihs {\sl et al.}.
The former is less
sophisticated in the sense that each station only has one detector
and has no facility to rapidly switch the orientation of the polarizer
in front of the detector~\cite{AGUE09}.
Therefore, for each fixed setting of the polarizers,
the experiment only produces a set of the time tag data for the
$x_{n,i}=+1$ (say) detection events for fixed settings $a$ and $b$.
The set of data collected at station $i$ may be written as
\begin{eqnarray}
\label{Ups2}
\widetilde\Upsilon_i=\left\{ {t_{n,i},\theta_{i} \vert n =1,\ldots ,N_i } \right\}
,
\end{eqnarray}
where $\theta_i=a,b$ for station 1 and 2, respectively.
Thus, for each particular choice of the settings $\theta_1$ and $\theta_2$, this experiment
only yields $C_{++}(\theta_1,\theta_2)$ but the counts for the events $x_{n,i}=-1$
and coincidences $C_{+-}(\theta_1,\theta_2)$, $C_{-+}(\theta_1,\theta_2)$,
and $C_{--}(\theta_1,\theta_2)$ can be obtained by rotating
the corresponding polarizer by 90 degrees and repeating the experiment.

Table~\ref{table.tab4} presents results of the analysis of data sets
provided to us by A.A. Hnilo.
The histogram of time-tag differences (not shown) shows that
essentially all coincidences fall within a time window of $W=25\ns$.
The total number of coincidences ($N_c=C_{++}+C_{--}+C_{-+}+C_{+-}$)
is larger than 9600 for all cases.
For $a=0$, $a'=\pi/4$, $b=\pi/8$ and $b'=3\pi/8$,
the data of Table~\ref{table.tab4} yields $|S(a,a',b,b')|=2.73$.
For these values of $a$, $a'$, $b$, and $b'$,
the differences between
$E_1(a,b)=0.129$ and $E_1(a,b')=0.087$,
$E_1(a',b)=0.033$ and $E_1(a',b')=-0.025$,
and
$E_2(a,b)=0.127$ and $E_2(a',b)=0.082$,
are all larger than 4 standard deviations
whereas the difference between
$E_2(a,b')=-0.059$ and $E_2(a',b')=-0.077$
is less than 2 standard deviations.

The experiment by M.M. Ag\"uero {\sl et al.} passes our hypothesis test
with a narrow margin, but nevertheless, deviations
by more than four standard deviations do not favor the
hypothesis that this data can be described by quantum
theory of the EPRB experiment
but may suggest that the systematic violation
of the five-standard deviation criterion by the data
of the experiment of Weihs {\sl et al.} may be due to
intricacies of the much more complicated experimental set up of the latter.

\begin{table}
\begin{center}
\caption{%
The coincidences for various settings $a$ and $b$ (both in degrees) of the polarizers,
extracted from the data sets provided to us by A.A. Hnilo~\cite{AGUE09}.
The time window $W=25\ns$ (the results for $W=12.5\ns$ are nearly identical),
and the global time shift $\Delta_G=0$.
In each run of the experiment, station 1 and station 2 detected about 90000 photons each.
The single detection count rate per second is about 500.
}
\smallskip
\label{table.tab4}       
\begin{tabular}{r|rr|rr|rr|rr}
\hline
$b$  & $a$  & $C_{++}(a,b)$ & $a$  & $C_{++}(a,b)$ & $a$  & $C_{++}(a,b)$
& $a$  & $C_{++}(a,b)$
\\
\hline
   0.0 &     0 &  5068 &     45 &   2693 &   90 &     96 &  135 &   2584 \\
  22.5 &     0 &  4879 &     45 &   4394 &   90 &    798 &  135 &    844 \\
  45.0 &     0 &  2502 &     45 &   4819 &   90 &   2087 &  135 &    131 \\
  67.5 &     0 &   776 &     45 &   3797 &   90 &   3753 &  135 &    748 \\
  90.0 &     0 &    73 &     45 &   2239 &   90 &   4380 &  135 &   2042 \\
 112.5 &     0 &   806 &     45 &    611 &   90 &   3591 &  135 &   3837 \\
 135.0 &     0 &  2348 &     45 &    128 &   90 &   2098 &  135 &   4413 \\
 157.5 &     0 &  4453 &     45 &   1005 &   90 &    643 &  135 &   4300 \\
 180.0 &     0 &  5256 &     45 &   2676 &   90 &     81 &  135 &   2630 \\
\hline
\end{tabular}
\end{center}
\end{table}

\section{EPRB laboratory experiment III}\label{facts3}

The third set of data that we analyze has been provided to us by G. Adenier.
Although very different in implementation (different source and setup, see Refs.~\citen{ADEN12,VIST12})
from the EPRB experiment of Weihs {\sl et al.},
conceptually it is the same except that there is no facility to rapidly switch
the orientation of the half-wave plate
in front of the polarizing beam splitter~\cite{ADEN12,VIST12}.

The data set collected at station $i$ by a run of the experiment of Adenier {\sl et al.}
may be written as
\begin{eqnarray}
\label{Ups3}
\widehat\Upsilon_i=\left\{ x_{n,i},{t_{n,i},\theta_{i} \vert n =1,\ldots ,N_i } \right\}
,
\end{eqnarray}
where $\theta_i=a,b$ for station 1 and 2, respectively.
In practice, Adenier {\sl et al.} keep $a$ fixed and sweep $b$ from 0 to 360 degrees.
The method to analyze these data sets is identical to the one used to analyze the data of Weihs {\sl et al.}~\cite{WEIH00}

\begin{figure}[t]
\centering
\includegraphics[width=7.5cm]{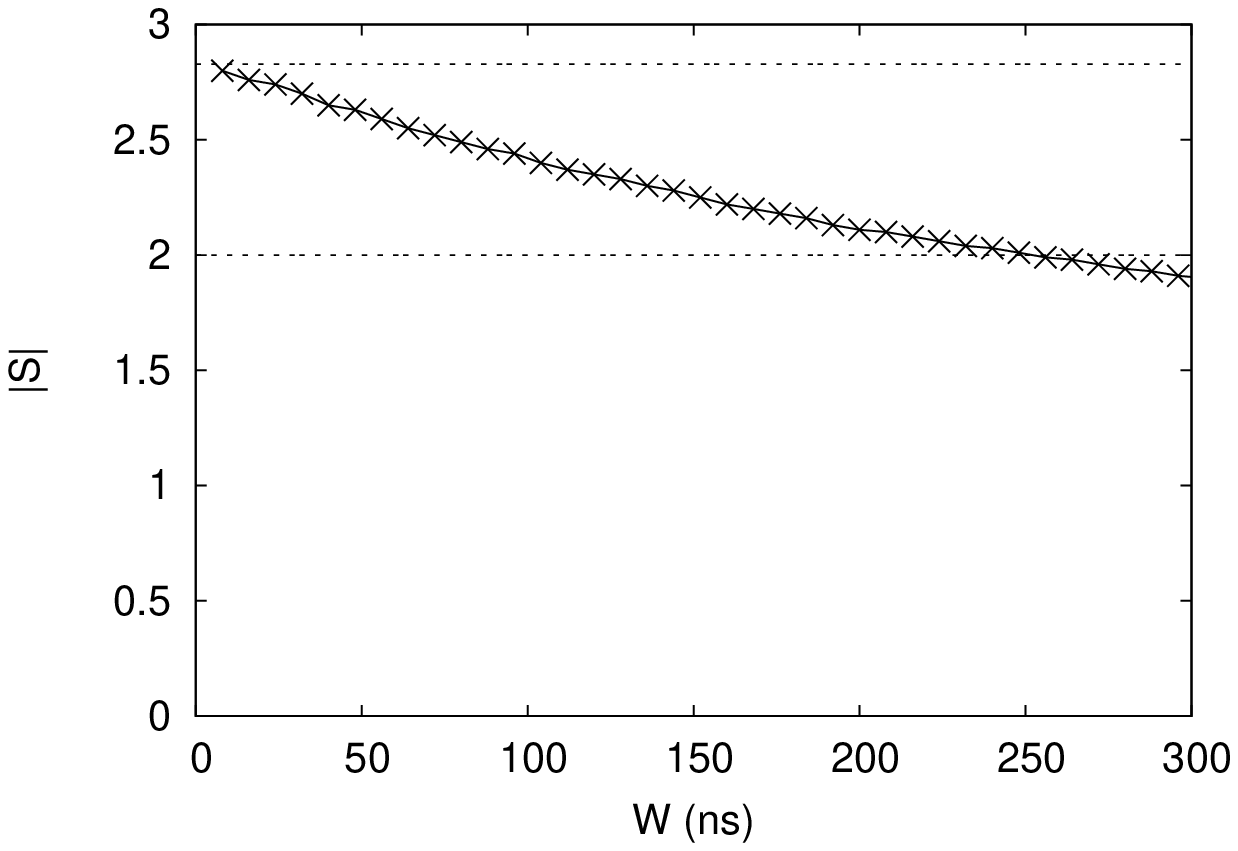}
\includegraphics[width=7.5cm]{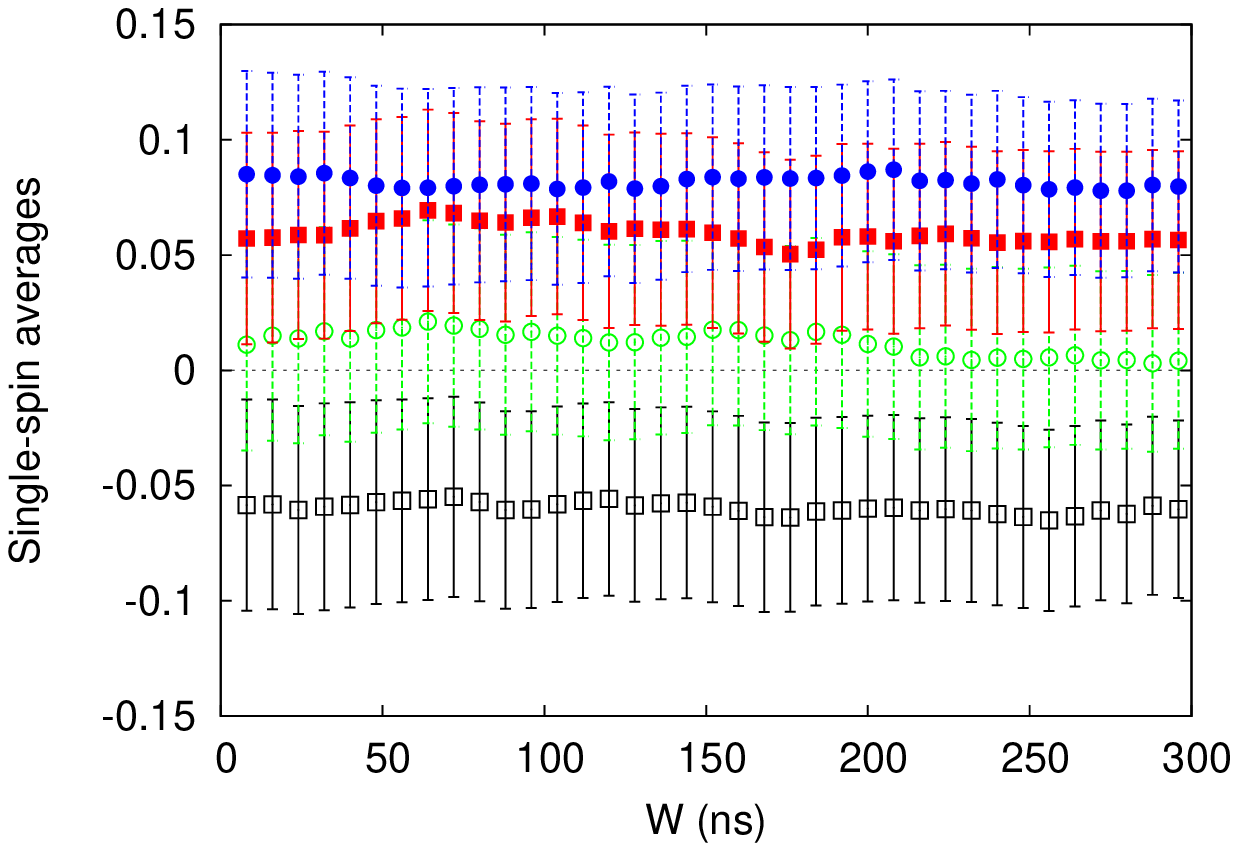}
\caption{
Analysis of the data sets provided to us by Adenier {\sl et al.}~\cite{ADEN12,VIST12}
Left:
$|S|=|E(a,b)-E(a,b')+E(a',b)+E(a',b')|$ as a function of the time window $W$ and
$a=0$, $a'=\pi/4$, $b=\pi/8$ and $b'=3\pi/8$.
The dashed lines represent the maximum value for a quantum system of two $S=1/2$ particles
in a separable (product) state ($|S|=2$) and in a singlet state ($|S|=2\sqrt{2}$), respectively.
Only if $|S|>2$, a Bell inequality is violated.
The line through the data points is a guide to the eye only.
Right:
Selected single-particle averages as a function of $W$.
Open squares: $E_1(a,b)$;
open circles: $E_1(a,b')$;
solid squares: $E_2(a,b)$;
solid circles: $E_2(a',b)$;
error bars correspond to $2.5$ standard deviations.
The total number of coincidences per setting pair is about 3000.
The change of  the single-spin averages $E_1(a,.)$
observed by Alice when Bob changes his setting ($b\rightarrow b'$) ,
is close to five standard deviations,
suggesting it is highly unlikely that the data is in
concert with quantum theory of the EPRB experiment.
}
\label{fig.C5}
\end{figure}

\begin{figure}[t]
\centering
\includegraphics[width=7.5cm]{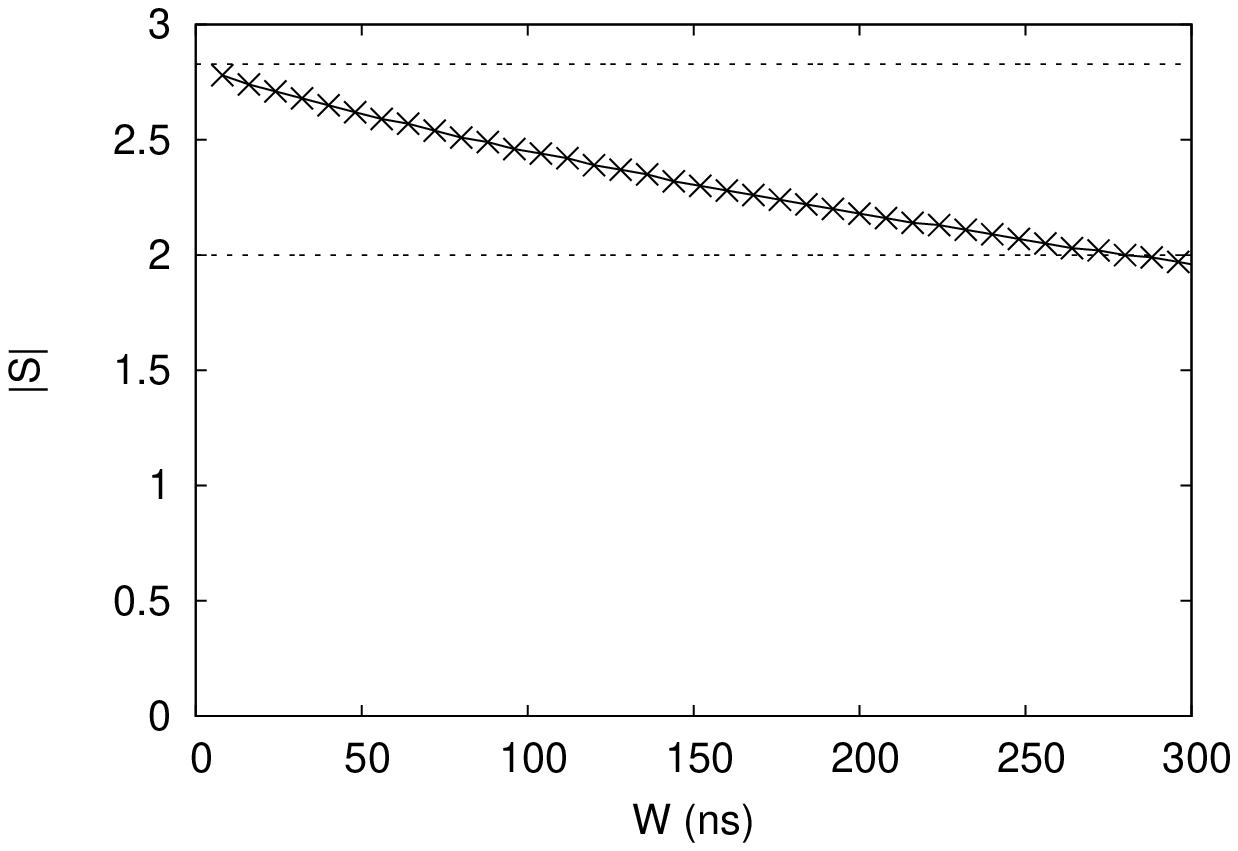}
\includegraphics[width=7.5cm]{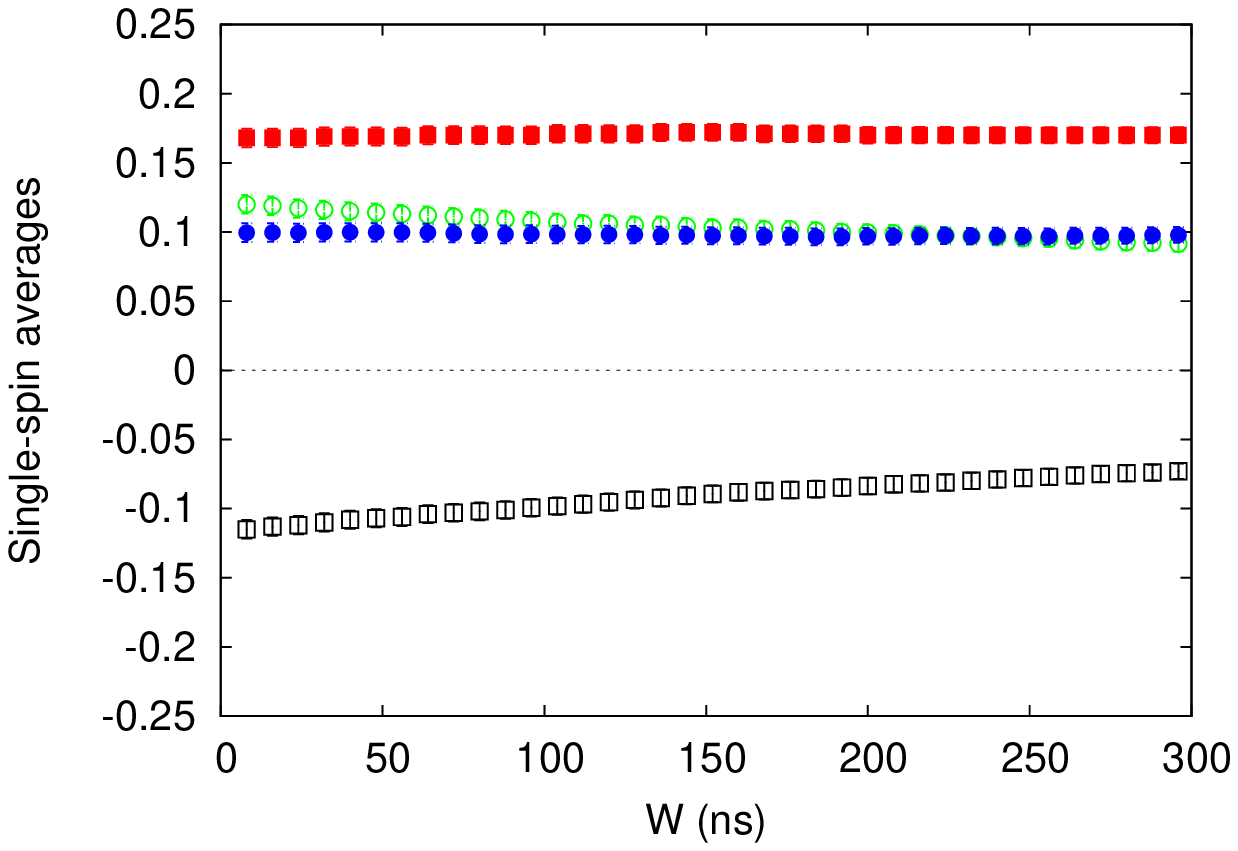}
\caption{
Same as Fig.~\ref{fig.C5} except that
the total number of coincidences per setting pair is about 140000.
The 2.5 standard deviation error bars are about the size of the markers.
}
\label{fig.C6}
\end{figure}

Results of our analysis are presented in Fig.~\ref{fig.C5}.
Qualitatively, the results for the single particle averages look very similar to those of Fig.~\ref{fig.A5},
except that the error bars are larger.
This may partially be due to the fact that the number of coincidences accumulated
in the data set of Adenier {\sl et al.} is less than in the data set {\bf newlongtime2}
of Weihs {\sl et al.}.
However, if the underlying (unknown) mechanism that produces the data
would comply with quantum theory, one may expect that
the agreement with quantum theory improves if the number of coincidences increases.
Remarkably, as shown by the data presented in Fig.~\ref{fig.C6},
this expectation is in contradiction with the experiment.
In conclusion, the data produced in the experiment of Adenier {\sl et al.}
does not pass our hypothesis either, hence as for the two previous experiments,
it is very unlikely that also this third experiment
can be described by quantum theory of the EPRB thought experiment.

\section{Discrete event simulation}\label{model}\label{simulationmodel}

The fact that it is difficult to reconcile the data produced by the three EPRB experiments
that we have analyzed with the predictions of quantum theory for the EPRB thought experiment
raises the question whether there exist
(simulation) models of this experiment that produce the same kind of data sets as the real experiment and
do not rely on concepts of quantum theory, yet reproduce the results of quantum theory of the EPRB experiment.

The possibility that such models exist was, to our knowledge, first pointed out by A. Fine~\cite{FINE82}.
The key of Fine's so-called synchronization model is the use of a filtering mechanism, essentially
the time-coincidence window employed in laboratory EPRB experiments.
Fine pointed out that such a filtering mechanism may lead to violations
of the inequality $|S|\le2$, opening the route to a description in terms of locally causal, classical models.
A concrete model of this kind was proposed by S. Pascazio who showed
that his model approximately reproduces the correlation of the singlet state~\cite{PASC86}
with an accuracy that seems far beyond what is experimentally accessible to date.
Later, Larson and Gill showed that Bell-like inequalities need to be modified
in the case that the coincidences are determined by a time-window filter~\cite{LARS04}.
Time-tag models that exactly reproduce the results of quantum theory for the singlet
and uncorrelated state were reported in Ref.~\citen{RAED06c,RAED07b,ZHAO08,MICH11a,RAED12a}.
These event-based simulation models provide a
cause-and-effect description of real EPRB experiments at a level of detail which is not covered
by quantum theory, such as the effect of the choice of the time-window.
Some of these simulation models exactly reproduce the results of quantum theory of the EPRB experiment,
indicating that there is no fundamental obstacle for an EPRB experiment
to produce data that can be described by quantum theory.

\section{Conclusion}\label{conclusion}

We have shown that it is highly unlikely that quantum theory describes
the data of three very different realizations of EPRB experiments,
independent of assumptions about the nature of the state of the two-particle quantum system
if this state does not depend on the setting of the analyzers.
Of course, this conclusion does not suggest that experiments have proven quantum theory wrong.
Our analysis does not rule out that the data
can be described by a state for which the state explicitly depends on
the settings $\mathbf{a}$ or $\mathbf{b}$, in which case the very essence of the EPRB experiment is lost.
It merely suggests that in the real experiments,
there are processes at work which have not been identified yet and
that better experiments are called for.

Our conclusion does not critically depend on the choice of the time-coincidence window $W$ that is used
in the experiments. Thereby it should be noted that the average time between two detector clicks is of the order of
$30\mus$, 2 ms, and $50\mus$ for experiment 1, 2, and 3, respectively
and that the values of $W$ for which $|S|$ drops below 2 is typically an order of magnitude smaller.
Therefore, the argument that the coincidence windows merely serves to ``identify'' pairs may be
too naive, as also indicated by the fact that event-based simulation models are able to produce results
which are in concert with quantum theory of the EPRB thought experiment.

Summarizing: the popular statement that EPRB experiments agree with quantum theory of the EPRB thought experiment
does not seem to have a solid scientific basis yet.

Future experiments which aim to establish that the EPRB experiment complies with quantum theory
should present a complete analysis of the data.
This should at least entail
\begin{enumerate}
\item
Plots such as Fig.~\ref{fig.A5}(left), showing the Bell function $S$ as a function of the time-coincidence window $W$.
Coincidences should be computed without ``compensating'' for a global time shift $\Delta_G$
to avoid introducing acausal effects by hand.
\item
Plots of the data for the single counts, similar to Fig.~\ref{fig.A5}(right), which display the data
as a function of the time-coincidence window $W$, for different pairs of setting such
($E_1(a,b),E(a,b')$). Such plots would reveal immediately whether there is a
mismatch with the predictions of quantum theory, which predicts $E_1(a,b)=E(a,b')$
and $E_2(a,b)=E(a',b)$ for any set of $a$, $a'$, $b$, and $b'$.
\item
Plots of the coincidence counts $C_{xy}(a,b)$.
\item
Plots of the two-spin correlation $E(a,b)$.
\end{enumerate}
Furthermore, as this paper shows, it is essential that the data is made publicly available
such that it can be tested against various hypotheses.

\section{Acknowledgment}
We thank G. Weihs, A.A. Hnilo and G. Adenier for providing us with the data sets of their EPRB experiments.
We have profited from discussions with G. Adenier, J.H. Bigelow, K. De Raedt, K. Hess, A. Khrennikov, S. Reuschel, A-I. Vistnes,
and G. Weihs.
This work is partially supported by NCF, the Netherlands.

\bibliographystyle{spiebib}   
\bibliography{/d/papers/all13}   

\end{document}